\documentclass[onecollarge]{svjour2}
\usepackage{graphicx,amssymb,amsmath}
\newcommand{\eg}{\textit{e.g.}}  

\journalname{Few-Body Systems}

\begin{document}


\title{
Electron Anomalous Magnetic Moment in Basis Light-Front Quantization Approach}

\author{ X.~Zhao, H.~Honkanen, P.~Maris, J.~P.~Vary, S.~J.~Brodsky}
\institute{X.~Zhao, P.~Maris, J.~P.~Vary \at
Department of Physics and Astronomy, Iowa State University, Ames, Iowa 50011, USA, \email{xbzhao, pmaris, jvary@iastate.edu}
          \and
           H.~Honkanen \at
         Department of Physics, University of Jyv\"askyl\"a, P.O. Box 35 (YFL), FI-40014 University of Jyv\"askyl\"a, Finland, \email{heli.m.honkanen@phys.jyu.fi}
          \and
          S.~J.~Brodsky \at
         SLAC National Accelerator Laboratory, Stanford University, Menlo Park, California, USA, \email{sjbth@slac.stanford.edu}
}
\date{\today}
\maketitle 
\begin{abstract}
We apply the Basis Light-Front Quantization (BLFQ) approach to the Hamiltonian field theory of Quantum Electrodynamics (QED) in free space. We solve for the mass eigenstates corresponding to an electron interacting with a single photon in light-front gauge. Based on the resulting non-perturbative ground state light-front amplitude we evaluate the electron anomalous magnetic moment. The numerical results from extrapolating to the infinite basis limit reproduce the perturbative Schwinger result with relative deviation less than 0.6\%. We report significant improvements over previous works including the development of analytic methods for evaluating the vertex matrix elements of QED.
\end{abstract}

\section{Introduction}
\label{intro}
Hamiltonian light-front theory has been proposed as a promising method for solving bound states problems in quantum field theory a long time ago, see, \eg, the review article\cite{Brodsky:1997de} for its various advantages compared to the traditional instant-form theories. Recently the Basis Light-Front Quantization (BLFQ) approach~\cite{Vary:2009gt,Honkanen:2010rc} has been developed as a nonperturbative approach to solve Hamiltonian light-front quantum field theory. Numerical efficiency is a key advantage of this approach. The basic idea of BLFQ is to represent the theory in an optimal basis which respects many symmetries of the theory and thus minimizes the dimensionality of the Hamiltonian for a fixed precision. Specifically, the BLFQ approach employs a plane wave basis in the light-front longitudinal direction and a 2D harmonic oscillator basis in the transverse directions. In previous work~\cite{Honkanen:2010rc} this approach has been applied to evaluate the anomalous magnetic moment of electrons which are confined in an external trap with an extrapolation to the zero trap limit. In this work we extend and improve this approach in several aspects including the direct evaluation of a free electron system. This article is organized as follows: In Sec.~\ref{prog} we discuss the key extensions and improvements made in this work over Ref~\cite{Honkanen:2010rc}; in Sec.~\ref{result} we present the numerical results for the electron anomalous magnetic moment evaluated in different harmonic oscillator bases and compare to the perturbation theory result. Finally we conclude and give an outline for future works in Sec.~\ref{end}.

\section{Extensions and Improvements}
\label{prog}
As in the previous work~\cite{Honkanen:2010rc} we apply BLFQ to QED and solve for the mass eigenstates for single electrons in a truncated basis including only $|e\rangle$ and $|e\gamma\rangle$ sectors. In this paper we briefly report the major extensions and improvements made over Ref.~\cite{Honkanen:2010rc}. A more systematic description for applying the approach to QED will follow in a separate paper~\cite{Zhao:2011}. Before introducing the details of these extensions and improvements, we list them here.\\

1. We extend the application of BLFQ to a free space system by omitting the external trap.\\

2. We replace numerical integrations previously used to evaluate matrix elements of interaction vertices in the QED Hamiltonian with newly-developed analytic methods.\\

3. We correct the evaluation of the anomalous magnetic moment from the electron light-front amplitude and a factor in the vertex matrix elements.  These corrections go in opposing directions for the
external trap results for the anomalous magnetic moment in Ref.~\cite{Honkanen:2010rc}.  The results of Ref.~\cite{Honkanen:2010rc} with an external trap will be fully updated in a future publication.\\

In the previous work~\cite{Honkanen:2010rc}, the system (a physical electron consisting of a constituent bare electron and a bare photon) was configured with an external trap provided by a scalar harmonic oscillator (HO) potential acting on both the (constituent) electron and photon in the transverse directions. The frequency 
$\omega$ of the HO basis was chosen to coincide with the frequency of the HO trap potential. The motivation for introducing this potential is two-fold: a) it provides an ultraviolet cutoff for the QED interactions; b) it simplifies the Hamiltonian matrix since the basis is made of the eigenstates of the noninteracting Hamiltonian. However this external potential may be difficult to implement experimentally due to its scalar nature and the fact that it confines both electrons and photons in the same trap. Therefore in order to compute quantities which are directly comparable with trap-free experimental data, it is desirable to set up the BLFQ approach in free space. To achieve this goal we implement the following procedures in the order we now list.\\

1) We decouple the HO frequency $\omega$ chosen for the basis from that of the trap potential. One of the limitations in Ref.~\cite{Honkanen:2010rc} is that the $\omega$ for the external trap potential was taken to be the same as that for the HO basis. However, in order to calculate observables in free space the $\omega$ for the trap potential needs to be taken to zero, while the $\omega$ for the HO basis has to remain finite. Note that, in principle, the basis is arbitrary as long as it is complete. Thus arbitrary $\omega$'s for the HO basis should lead to identical physical observables. However, different choices of 
$\omega$ will exhibit different rates of convergence with increasing basis dimension to the ``true" value, due to different effective infrared and ultraviolet cutoffs associated with each $\omega$.  Furthermore, the rates of convergence will also depend on the specific observable evaluated. We will illustrate this point for the anomalous magnetic moment in Sec.~\ref{result}. Decoupling these two $\omega$'s increases the number of nonzero off-diagonal matrix elements of the Hamiltonian matrix, but only mildly, because the kinetic energy term in the Hamiltonian, $\sim p^2$, in the HO basis only couples states with adjacent principle quantum numbers leading to tridiagonal kinetic components in the Hamiltonian.\\

2) We remove the transverse center-of-mass (cm) kinetic energy from the Hamiltonian. For the free space states only the relative transverse motion of the constituent particles is relevant for its invariant mass spectrum and light-front amplitudes. However, since BLFQ is set up in the single-particle coordinates, the kinetic energy of the transverse cm of all the constituent particles is naturally included in the sum over kinetic energies for all particles. Therefore the explicit removal of the transverse cm kinetic energy from the Hamiltonian is necessary. We achieve this by making the following replacement for the transverse momenta of all constituent particles $n$ in a given Fock sector: $p_i\to p'_i=p_i-\frac{\sum_i{p_i}}{n}$. It is easy to verify that $\sum_i p'_i=0$ and therefore the $p'_i$ carries only the relative transverse momentum. \\

3) We remove all the excited configurations and keep only the lowest configuration in the single-electron sector, $|e\rangle$. With the cm kinetic energy removed from the Hamiltonian, the basis is overcomplete for representing the internal motion among the constituent particles. This overcompleteness may lead to multiple copies of the mass eigenstates with identical internal motion but different cm motions, which cause unnecessary numerical burden and complicate the mass spectrum. Therefore, it is desirable to remove these (spurious) copies from the spectrum. In this work we achieve this by simply removing all the excited configurations in the single-electron sector. Since these configurations produce states with excited states of cm motion, the removal of these configurations does not affect the completeness of the basis for representing the internal motion of constituent particles through higher Fock sectors. Furthermore, since the single-electron sector is the dominant sector for (physical) electrons in QED, removing the (spurious) copies of the cm motion from the single-electron sector effectively eliminates cm excitations from the lowest mass eigenstates. That is, the Hamiltonian of QED conserves cm motion and all excited states of cm motion will now involve only the $|e\gamma\rangle$ configurations not coupled by QED to the $|e\rangle$ configuration that dominates the lowest mass eigenstate.\\

4) We implement sector-dependent HO basis frequencies ($\omega$'s). Different Fock sectors are coupled by the interaction terms in the QED Hamiltonian. These interaction terms are identity operators with respect to the transverse cm motion. Now, in the single-electron sector we already chose the cm motion to be in the lowest allowed HO state. Thus, it is desirable to adopt the same state of cm motion for all higher Fock sectors. For Fock sectors with $n$ particles, the cm motion can be described by HO bases with the characteristic length parameter $b_n\equiv \sqrt{nM\omega_n}$, where $M$ is the mass of the HO specifying the basis. (Note that this $M$ has nothing to do with the physical mass of particles in the QED Hamiltonian and, for convenience, all single-particle basis states in all Fock sectors are assumed to have the same $M$). In order to match $b_n$ for all Fock sectors, the HO frequency in each $n$-particle Fock sector is required to satisfy $\omega_n$=$\frac{1}{n}\omega_1$, where $\omega_1$ is the basis frequency for the single electron sector. \\

The second improvement is related to the calculation of the matrix elements of the interaction vertices between different BLFQ states. Let us take the matrix elements coupling $|e\rangle$ and $|e\gamma\rangle$ sectors for example. To evaluate these matrix elements one needs to perform integrations in the transverse momentum space over a product of three HO wave functions associated with the $e$ in $|e\rangle$ and the $e$,$\gamma$ in $|e\gamma\rangle$ sectors. In Ref.~\cite{Honkanen:2010rc} these integrations were computed numerically in the single-particle representation. Due to the oscillatory structure of HO wave functions, the numerical computation was very time-consuming to achieve results with the desired precision. In this work we reduce the numerical integrations by applying an analytical Talmi-Moshinsky transform \cite{Talmi-Moshinsky} to simplify the integrations. The needed 2D Talmi-Moshinsky coefficients are analytically worked out in Ref.~\cite{Chaos-Cador:2004}. 

Adopting this analytic step greatly reduces the calculation time for the desired precision of the resulting matrix elements. For example, numerically integrating all matrix elements coupling $|e\rangle$ and $|e\gamma\rangle$ sectors with the truncation at $N_{\rm{max}}$=40\footnote{In BLFQ $N_{\rm{max}}$ indicates the upper limit for total HO quantum numbers summed over all particles in a Fock sector. Any state whose total HO quantum number exceeds $N_{\rm{max}}$ are omitted. For $N_{\rm{max}}$=40 there are 8190 unique non-vanishing electron to electron-photon vertex matrix elements in total.} previously took $\sim$7000 cpu-hours at the National Energy Research Supercomputer Center (NERSC) and only $\sim$15\% of them have relative errors better than one part in 10$^{7}$.  With the improved methods, we now evaluate these same matrix elements in less than 2 cpu-seconds and all of them have relative errors better than one part in 10$^{7}$.  Developing and adopting these analytic techniques for interaction matrix elements solves a bottleneck in BLFQ and greatly expands the attainable basis space dimensions. Increasing basis space dimension is crucial for evaluating observables with slow convergence with $N_{\rm{max}}$ and/or improving the overall precision. We expect that the same analytic techniques can be applied to calculate all interacting vertices in QED and QCD. The details for the analytic calculation of the interaction matrix elements will be given in a forthcoming paper~\cite{Zhao:2011}.

Our third improvement is in the method of extracting the electron anomalous magnetic moment from its light-front amplitude obtained from diagonalization of the QED Hamiltonian matrix. We also correct a factor in the calculation of the vertex matrix elements in the Hamiltonian matrix.  These two corrections to the results in Ref.~\cite{Honkanen:2010rc} have approximately offsetting consequences.  

In Ref.~\cite{Honkanen:2010rc} the electron anomalous magnetic moment was approximated (based on non-relativistic quantum mechanics) by the squared modulus of the helicity-flip (for the constituent electron) components of the eigenstates. The precise definition of electron anomalous magnetic moment in relativistic QED is $a_e$, the electron Pauli form factor $F_2$ evaluated at momentum transfer $q^2\to 0$~\cite{Brodsky:1980zm},
\begin{align}
\label{g-2}
a_e\equiv\frac{g-2}{2}=F_2(q^2\to0) .
\end{align}
In BLFQ the $a_e$ can be calculated by sandwiching the operator $F_2(0)$ with the obtained ground states for the electron,
\begin{align}
\label{g-2_blfq}
\nonumber a_e&\equiv\frac{g-2}{2}=\langle\Psi^{\downarrow}_e|F_2(0)|\Psi^{\uparrow}_e\rangle\\
&=\sum_{i',i}\langle\Psi^\downarrow_e|e\gamma,i'\rangle\langle e\gamma,i'|F_2(0)|e\gamma,i\rangle\langle e\gamma,i|\Psi^\uparrow_e\rangle .
\end{align}
Here $\langle e\gamma,i'|F_2(0)|e\gamma,i\rangle$ is the matrix element of the Pauli form factor in the BLFQ basis. The $\langle e\gamma,i|\Psi^{\uparrow(\downarrow)}_e\rangle $ is the wavefunction of a physical electron with helicity up (down) in the $|e\gamma\rangle$ sector (the only sector contributing to $a_e$ in our truncated basis). The $i$ denotes a complete set of quantum numbers. Although Eq.(\ref{g-2_blfq}) involves two electron eigenstates with opposite helicities, in practice one needs only to solve for one of them and infer the other by exploiting the parity symmetry in light-front QED~\cite{Brodsky:2006ez}. The explicit expression for $\langle e\gamma,i'|F_2(0)|e\gamma,i\rangle$ and the exact relation between $\langle e\gamma,i|\Psi^\uparrow_e\rangle$ and $\langle e\gamma,i|\Psi^\downarrow_e\rangle$ will be reported in a later work~\cite{Zhao:2011}.

As noted above, we also correct the interaction matrix elements used in Ref.~\cite{Honkanen:2010rc} where there was an inadvertent omission of a factor of 2. This, in turn led to an overestimate of the evaluated square root of anomalous magnetic moment based on the spin-flip modulus by a factor of 2. Fortuitously, according to an analysis based on the perturbation theory~\cite{Zhao:2011}, the modulus of the helicity flip amplitude is down by a factor of $\sqrt{2}$ from $a_e$ defined above. Therefore, these two corrections are partially offsetting. Further improvements to the methods of Ref.~\cite{Honkanen:2010rc} for the extrapolation to the free space limit are now envisioned based on the sensitivities we observe in the results presented here.  The revised results for an electron in an external trap will be presented in a later work~\cite{Zhao:2011} and will include results in significantly larger basis spaces to underpin an improved extrapolation to the free-space limit.

\section{Numerical Results for Electron $a_e$}
\label{result}
Incorporating all the improvements introduced in Sec.~\ref{prog}, we diagonalize the light-front QED Hamiltonian in $|e\rangle$ and $|e\gamma\rangle$ sectors and extract the anomalous magnetic moment from the resulting light-front amplitude for the lowest mass eigenstate. In this work we reduce the QED coupling constant $\alpha$ by a factor of 10$^7$ to limit the difference between our non-perturbative result and the perturbative result. Then, in this small coupling limit, the contributions from higher order perturbations can be neglected so that the Schwinger result~\cite{Schwinger:1948iu} can be used as the reference for our numerical results. 

We evaluate and diagonalize the Hamiltonian with total longitudinal momentum (summed over the constituent electron and photon) $K$=160, translating to 160 (equal distance) partitions between the longitudinal momentum of the constituent electron and  photon. In the transverse directions we use HO basis with several selected frequencies $\omega$'s, ranging from 0.01MeV to 1.4MeV bracketing the electron mass $m_e$=0.511MeV, the only dimensionful quantity in the QED Hamiltonian. At each $\omega$ we calculate a series of results with increasing $N_{\rm{max}}$ from 10 to 118, for the purpose of mapping out the convergence behavior of anomalous magnetic moment with respect to $N_{\rm{max}}$. The larger $N_{\rm{max}}$ translates to a larger basis (the largest here consists of 44,358,561 basis states) with higher effective ultraviolet cutoff and lower effective infrared cutoff in the transverse plane. It is thus expected that, with increasing $N_{\rm{max}}$, the results grow closer to the Schwinger result. The rate of  convergence may, however, be different for different $\omega$'s, depending on the sensitivity of the anomalous magnetic moment to the infrared and ultraviolet cutoffs in the transverse directions. Indeed our numerical results agree with this expectation: all the results with $\omega$ between 0.01MeV to 1.4MeV are below the Schwinger result at finite $N_{\rm{max}}$ which they approach uniformly as $N_{\rm{max}}$ increases with increments of 4.

\begin{figure}[!t]
\centering
\includegraphics[width=0.8\textwidth]{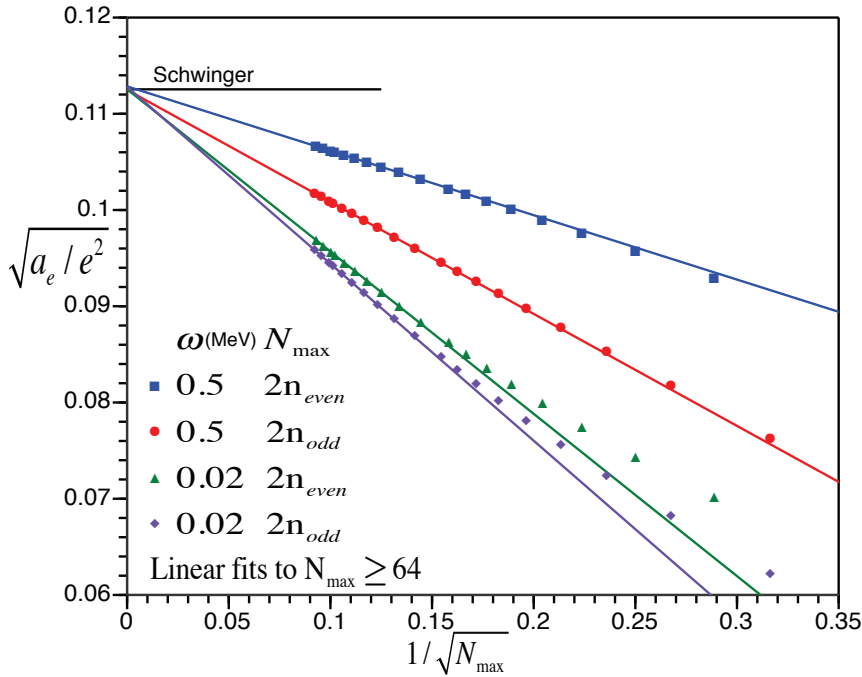}
\caption{(Color online) Anomalous magnetic moment of the electron calculated in BLFQ compared to the Schwinger result~\cite{Schwinger:1948iu}. The vertical axis is the square root of anomalous magnetic moment normalized to electron charge, $e$, so the Schwinger value is $\sqrt{\frac{1}{8\pi^2}}=0.11254$. The horizontal axis is the square root of the reciprocal of $N_{\rm{max}}$. Symbols are for the BLFQ results. Squares: $\omega$=0.5MeV at even $N_{\rm{max}}$/2; circles: $\omega$=0.5MeV at odd $N_{\rm{max}}$/2; triangles: $\omega$=0.02MeV at even $N_{\rm{max}}$/2; diamonds: $\omega$=0.02MeV at odd $N_{\rm{max}}$/2. The lines are linear extrapolations of BLFQ results based on the points at or above $N_{\rm{max}}=64$.}
\label{anoma}
\end{figure}

In Fig.~\ref{anoma} we compare the results evaluated with $\omega$=0.02MeV and $\omega$=0.5MeV.
At the same $N_{\rm{max}}$ the results at $\omega$=0.5MeV are closer to the Schwinger result than those at $\omega$=0.02MeV. For each $\omega$ the results exhibit a distinctive pattern with increasing $N_{\rm{max}}$: the results with even $N_{\rm{max}}/2$ are systematically larger than those with odd $N_{\rm{max}}/2$ so that the former and the latter separate into two individual groups. Within each group the results define a simple trend. 

The systematics shown in Fig.~\ref{anoma} have simple explanations.  According to perturbation theory~\cite{Brodsky:1980zm}, the $a_e$ receives its largest contributions from the infrared region of the relative momentum between the constituent $e$ and $\gamma$. Thus, a difference between neighboring even and odd $N_{\rm{max}}$ points originates from a truncation-induced distortion of the (point-like) contact interactions (coupling $|e\rangle$ and $|e\gamma\rangle$ sectors) in the infrared region of momentum space. If one works in the relative coordinates (for the constituent $e$ and $\gamma$), representing such (point-like) contact potentials in a truncated HO basis typically leads to an alternating over- and under-estimate in the infrared region of the relative momentum at even (odd) $N_{\rm{max}}/2$ truncations. It is natural to expect that such qualitative features are the same in the single-particle coordinates used in BLFQ. 

The $\omega$ dependence of the convergence is also understandable. Since the HO basis with smaller $\omega$'s have more accurate representation for the infrared region, the odd-even oscillation of results at $\omega$=0.02MeV is milder than those at $\omega$=0.5MeV. On the other hand, the results at $\omega$=0.5MeV are closer to convergence at any chosen $N_{\rm{max}}$ reflecting naturalness of this basis for the QED problem with only the electron mass scale. 

For sufficiently large $N_{\rm{max}}$ the data points in Fig.~\ref{anoma} define straight lines as a function of $1/\sqrt{N_{\rm{max}}}$. We can therefore easily extrapolate to the limit of without basis truncation  ($N_{\rm{max}}\to\infty$) where we expect to recover the Schwinger result. Indeed as seen in Fig.~\ref{anoma} the four extrapolating lines converge close to the Schwinger result in this limit. Their intercepts at 1/$N_{\rm{max}}$=0 are: 0.11285(0.5\%), 0.11248(0.1\%), 0.11255(0.02\%) and 0.11280(0.5\%) for $\omega$=0.5MeV at even $N_{\rm{max}}/2$, $\omega$=0.5MeV at odd $N_{\rm{max}}/2$, $\omega$=0.02MeV at even $N_{\rm{max}}/2$, $\omega$=0.02MeV at odd $N_{\rm{max}}/2$, respectively. The percentages in the parenthesis are their corresponding relative deviation from the Schwinger result, $\frac{a_e}{e^2}=\frac{\alpha}{2\pi e^2}$=$\frac{1}{8\pi^2}\approx$0.012665.  These results suggest that the $\omega$'s between 0.02MeV and 0.5MeV form an ``optimal'' plateau, where the Schwinger result is reproduced with less than 0.6\% deviation. We also note the large width of this plateau (from $\omega$=0.5MeV to 0.02MeV amounting to a factor of 25), illustrating the flexibility and stability of the BLFQ approach. 
For $\omega$ slightly below 0.02MeV or above 0.5MeV reasonable precision can still be achieved and details will be presented in a later work~\cite{Zhao:2011}. 
Finally, we would like to emphasize the importance of large $N_{\rm{max}}$ results in extracting high precision $a_e$, since only at sufficiently large values of $N_{\rm{max}}$ do these points follow a straight line in $1/\sqrt{N_{\rm{max}}}$, allowing for reliable linear extrapolations as seen in Fig.~\ref{anoma}.

\section{Conclusions and Outlook}
\label{end}
Following Ref.~\cite{Honkanen:2010rc} we apply the BLFQ approach to QED and solve for the electron anomalous magnetic moment. Several extensions and improvements are introduced over the previous work~\cite{Honkanen:2010rc}. By removing the transverse external trap we are now able to solve for the mass eigenstates of QED in free space so that the resulting observables are directly comparable with most experimental data. The analytic procedures introduced for interaction matrix elements greatly reduce the total computing time and improve the overall precision. As a result, we can now access much larger basis spaces in BLFQ calculations, which is crucial for evaluating observables having slower convergence rates. As an application of BLFQ we evaluated the electron anomalous magnetic moment and compared to the Schwinger result. We found that the extrapolated values agree with the Schwinger result within reasonable precision with the present truncations.  We found that high precision results can be obtained with HO basis spaces over a wide range of frequencies, signifying the flexibility and stability of the BLFQ approach.

We plan further developments. 1) We will investigate renormalization schemes for light-front field theory such as those presented in Refs.~\cite{Brodsky:1973kb,Karmanov:2008br} and  calculate QED with the physical coupling. 2) We will extend the basis space to include higher Fock sectors and include all the interacting vertices in the QED Hamiltonian (including instantaneous vertices). These steps will allow BLFQ to capture all the main physics in QED and make it an ideal approach for investigating QED in nonperturbative regimes, such as in the presence of strong pulsed laser fields. 3) Furthermore, we will apply the BLFQ approach to QCD, where anticipate the BLFQ approach will be able to provide valuable information on the nonperturbative aspects of QCD. That is, in one perspective, we will extend the AdS/QCD approach of Ref.~\cite{deTeramond:2008ht} to include QCD interactions.

\vspace{5 mm}

The authors thank A. Harindranath, K.~Tuchin, J.~Hiller, S.~Chabysheva, 
V.~Karmanov, A.~Ilderton, Y. Li and P. Wiecki for fruitful discussions.  
Computational resources were provided by the National Energy Research Scientific Computing Center (NERSC), which is supported by the Office of Science of the U.S. Department of Energy under Contract No. DE-AC02-05CH11231.
This work was supported in part by US DOE Grants DE-FG02-87ER40371,  DE-FC02-09ER41582 (UNEDF SciDAC Collaboration) and by DOE Contract No. DE-AC02-76SF00515.
This work was also supported in part by US NSF grant 0904782.



\begin{thebibliography}{99}
%
%
%
%
%
%

\bibitem{Brodsky:1997de}
  S.~J.~Brodsky, H.~C.~Pauli and S.~S.~Pinsky,
  Phys.\ Rept.\  {\bf 301} (1998) 299.

\bibitem{Vary:2009gt}
J.~P.~Vary, H.~Honkanen, J.~Li, P.~Maris, S.~J.~Brodsky, A.~Harindranath, G.~F.~de Teramond, P.~Sternberg, E. G. Ng and C. Yang, 
  Phys.\ Rev.\  C {\bf 81},(2010) 035205. 

\bibitem{Honkanen:2010rc}
  H.~Honkanen, P.~Maris, J.~P.~Vary and S.~J.~Brodsky,
  Phys.\ Rev.\ Lett.\  {\bf 106}, 061603 (2011).

\bibitem{Zhao:2011}
  X.~Zhao, H.~Honkanen, P.~Maris, J.~P.~Vary and S.~J.~Brodsky,
  in preparation.

\bibitem{Brodsky:2006ez}
  S.~J.~Brodsky, S.~Gardner and D.~S.~Hwang,
  Phys.\ Rev.\  D {\bf 73}, 036007 (2006).

\bibitem{Brodsky:1980zm}
  S.~J.~Brodsky and S.~D.~Drell,
  Phys.\ Rev.\  D {\bf 22}, 2236 (1980).



\bibitem{Talmi-Moshinsky}
I.~Talmi. Helv. Phys. Acta, {\bf 25},185 (1952);
M.~Moshinsky, Nucl. Phys., {\bf 13}, 104(1959).

\bibitem{Chaos-Cador:2004}
  L.~Chaos-Cador, E.~Ley-Koo,
  Int.\ J.\ Quan.\ Chem.\  {\bf 97}, 844 (2004).


\bibitem{Schwinger:1948iu}
  J.~S.~Schwinger,
  Phys.\ Rev.\  {\bf 73} (1948) 416.

\bibitem{Brodsky:1973kb}
  S.~J.~Brodsky, R.~Roskies and R.~Suaya,
  Phys.\ Rev.\  D {\bf 8}, 4574 (1973).

\bibitem{Karmanov:2008br}
V.~A.~Karmanov, J.~F.~Mathiot and A.~V.~Smirnov,
Phys.\ Rev.\  D {\bf 77} (2008) 085028.

\bibitem{deTeramond:2008ht}
G.~F.~de Teramond and S.~J.~Brodsky,
Phys.\ Rev.\ Lett.\  {\bf 102}, 081601 (2009).


%
%
%
%
%
%
%
%
%





\end{thebibliography}
\end{document}